\documentclass[referee]{raa}            % referee version: for submission

\usepackage{graphicx,times}             %for PS/EPS graphics inclusion, new
\usepackage{natbib}
\bibpunct{(}{)}{;}{a}{}{,}

\usepackage[a4paper=true,dvipdfm=true,pagebackref=true]{hyperref}
\hypersetup{pdftitle = The title of my PDF, pdfauthor = My name, pdfsubject= The subject, pdfkeywords = keyword1 keyword2 keyword3}
\hypersetup{colorlinks = true, linkcolor = green, anchorcolor = red, citecolor = blue, filecolor = red, pagecolor = red, urlcolor = red}

\begin{document}

\title{Composing Method for the Two-dimensional Scanning Spectra Observed by the \emph{New Vacuum Solar Telescope}
\,$^*$
\footnotetext{$*$ Supported by the National Natural Science Foundation of China. }
\footnotetext{$\dagger$ Corresponding author}
}

\volnopage{Vol.0 (200x) No.0, 000--000}      %%preserved for Editor. DOn't remove!
   \setcounter{page}{1}          %%starting page, preserved for Editor. DOn't remove!

   \author{Yun-Fang Cai
      \inst{1,2}
   \and Zhi Xu
      \inst{1}\,$^\dagger$
   \and Yu-Chao Chen
      \inst{1,2}
   \and Jun Xu
      \inst{1}
   \and Zheng-Gang Li
      \inst{1}
   \and Yu Fu
      \inst{1}
   \and Kai-Fan Ji
      \inst{1}
   }

   \institute{Yunnan Astronomical Observatory, National Astronomical Observatories, Chinese Academy of Sciences,Kunming 650011, China; {\it xuzhi@ynao.ac.cn}\\
   \and
   University of Chinese Academy of Sciences, Beijing, 100049, China\\}

\date{Received~~2017 month day; accepted~~2017~~month day}

\abstract{In this paper we illustrate the technique used by the \emph{New Vacuum Solar Telescope} to increase the spatial resolution of two-dimensional (2D) solar spectroscopy observation involving two dimensions of space and one of wavelength. Without an image stabilizer at the NVST, a large scale wobble motion is present during the spatial scanning, whose instantaneous amplitude could reach up to 1.3 $''$ due to the earth's atmosphere and the precision of the telescope guiding system, and seriously decreases the spatial resolution of 2D spatial maps composed with the scanning spectra. We make the following effort to resolve this problem: the imaging system (e.g., the TiO-band) is used to record and detect the displacement vectors of solar image motion during the raster scan, in both the slit and scanning directions. The spectral data (e.g., the H$\alpha$ line) which are originally obtained in time sequence are corrected and re-arranged in space according to those displacement vectors. Raster scans are carried out in several active regions with different seeing conditions (two rasters are illustrated in this paper). Given a certain spatial sample and temporal resolution, the spatial resolution of the composed 2D map could be close to that of the slit-jaw image. The resulting quality after correction is quantitatively evaluated with two methods. Two-dimensional physical quantity, such as the line-of-sight velocities in multi-layer of the solar atmosphere, is also inferred demonstrating the effect of this technique.
}

   \authorrunning{Y.-F. Cai et al. }            %author_head in even pages
   \titlerunning{Composing Method for the Two-dimensional Scanning Spectra}  % title_head in odd pages
   \maketitle

%
%________________________________________________ sections below
%
\section{Introduction}           %% first-level sections will be auto-capitalized
\label{sect:intro}
Spectroscopy in two spatial dimensions (two dimensions of space and one of wavelength) is widely used by a large number of quantitative measurements of atmospheric structures in the field of solar physics that are retrieved only from the high resolution spectral observations, such as, the pressure, density, temperature and velocity etc. (\citealt{Fang1995}). Classical methods to acquire such 2D spectroscopy mainly include: narrow-band filtergrams, multi-channel subtractive double pass (MSDP, \citealt{1977Mein}) spectrographs and spatial scanning of slit spectrographs. Each kind of them has specific advantages and drawbacks. With the narrow-band filtergram, for example, the Fabry-Perot Interferometer (FPI, \citealt{Bonaccini1990}), monochromatic images are taken at successive wavelengths across a spectral line. A synthesis spectral line at each spatial position needs to be obtained within a certain time interval by this wavelength-scanning. The spectral resolution is in the order of 30 m\AA\ (e.g., the report by \citealt{Puschmann2012}). As contrast, the MSDP mode can provide monochromatic images at different wavelengths simultaneously. But its problem is that the observed field-of-view (FOV) is inversely with the spectral resolution, which means only a moderate spectral resolution can be achieved for a reasonable FOV. For instance, the MSDP installed in Meudon Solar Tower can provide the images at 9 wavelength points simultaneously with a FOV of 71 $''$ (perpendicular to the slit direction) $\times$ 465 $''$ (slit direction) and the wavelength distance is about 0.3 \AA \ (\citealt{2009Mein}). Spatial scanning is the method usually utilized together with a slit spectrograph. It enables one to use all the features of the spectrograph, such as high spectral (better than about 20 m\AA) and spatial resolution, wide wavelength range and simultaneous observations of multi-spectral lines. One can implement the telescope movement or equip a field scanner in front of the spectrograph to do spatial scanning, i.e., to move the solar image through the entrance slit continuously and rapidly. As known, the scanning speed should be the faster the better for a given solar active region since the solar dynamic structure continuously evolves with time. Compared with telescope drift, it is easy to use the field scanner to move the solar image fast and control the scanning speed according to the exposure time or other operations (e.g., the polarimetric modulation at each slit position for the spectro-polarimetry observation).

Usually for ground-based observations, it is not so easy to move the solar image rigidly in the right direction that is perpendicular to the entrance slit. The influence of both the earth's atmosphere and the precision of the telescope guiding system might cause the image wobble motion in both the slit and scanning directions. A 2D map directly composed by the spectral data, that are sequentially obtained with time during scanning, could has a very low spatial resolution (\citealt{Keller1990}). Solar structures even show discontinuities or zig-zag morphologies in this case. So in addition to the long time consumption mentioned above, another important thing involved in spatial scanning is how to guarantee the entrance slit located at a right position during scanning. Some efforts and techniques are made. For instance, \citealt{Johannesson1992} carry out the simultaneous acquisition of the spectral data and the slit-jaw images. A series slit-jaw images are used to monitor the image motion during scanning, allowing the spectrum to be re-mapped spatially during the data processing. Soon after, a solar correlation tracker is developed (\citealt{Ballesteros1996}), which can be used as an image stabilizer and an accurate positioning device to achieve 2D high spatial resolution spectra at real time (\citealt{Collados1996}). Nowadays, the adaptive optics (AO) system has been usually used by ground-based telescopes to resolve the problems of the image motion and distortion induced by the earth's atmosphere, such as, \emph{the German Vacuum Tower Telescope} (\citealt{2006Mikurda}), \emph{Gregory Coud¨¦ Telescope} (\citealt{Sutterlin2000}) and \emph{New Solar Telescope} (NST) (\citealt{Cao2010}), etc.

In this paper, we present a technique used by the \emph{New Vacuum Solar Telescope} (NVST) (\citealt{Liu2014}) in order to increase the spatial resolution of the 2D spectroscopy. It is similar to the method suggested by \citealt{Johannesson1992}, but the specific details are different. Meanwhile, the amplitude of the wobble motion is much larger in our case since there is no image stabilizer or AO system well working yet. We will introduce the data process step by step in the following sections.

The paper is organized as follows: Section ~\ref{S:1} introduces the characteristics of the instrument and observational setups. Section ~\ref{S:2} describes our technique in details. In Section ~\ref{S:3} we evaluate the composed image quality resulted from this technique. The conclusion and discussion are given in Section ~\ref{S:4}.

%%%%%%%%%%%%%%%%%%%%%%%%%%%%%%%%%%%%

\section{Instrumentations and Observations} %%%%%%%%%%%%%%%%%%%%%%%%%%%%%%%%%%%%%%%%
 \label{S:1}

\begin{figure}
\centering
\includegraphics[width=\textwidth,angle=0]{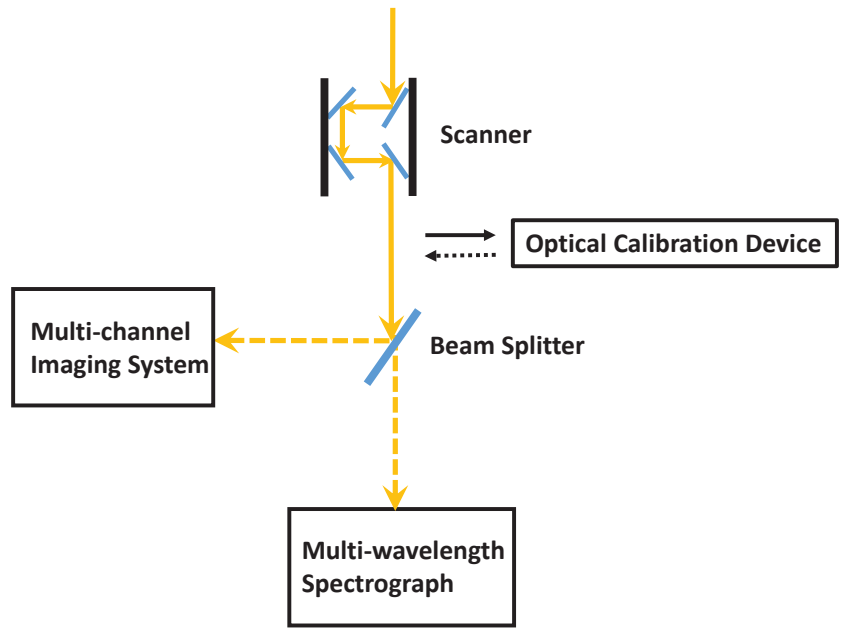}
\caption{Schematic of optical layout of the NVST instruments (Not to scale).}
\label{f:1}
\end{figure}

\begin{figure}
\centering
\includegraphics[width=\textwidth,angle=0]{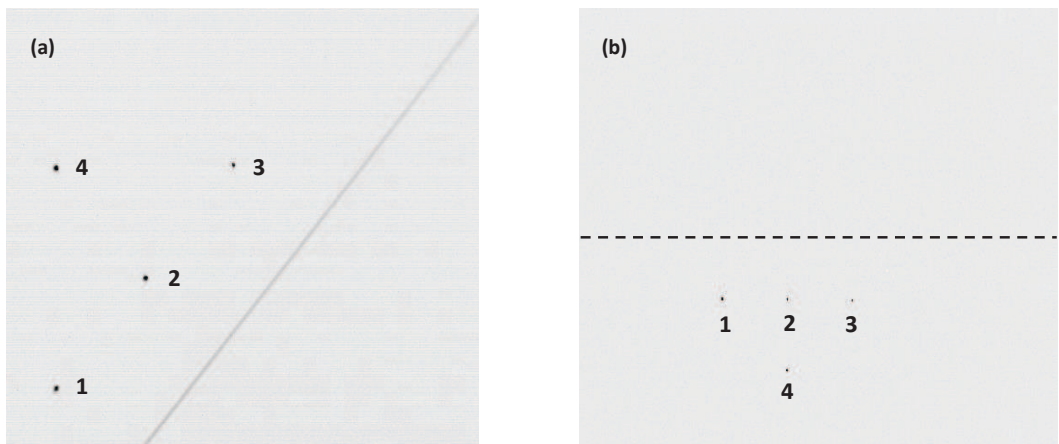}
\caption{Determination of the slit position in the FOV of the TiO image. (a) An image taken by a optical calibration device. Both an entrance slit of the spectrograph and a four-pin-hole diaphragm sitting at the focus plane of the imaging system are clearly seen. (b) An image of the four-pin-hole diaphragm taken at the TiO channel. The position of the entrance slit can be derived from the relative displacement shown in the panel (a) and denoted by a dashed line.}
\label{f:2}
\end{figure}

NVST is a vacuum solar telescope with a 985 mm clear aperture located at the Fuxian Solar Observatory of the Yunnan Observatories (\citealt{Liu2014}). Two instruments are operational up to now. One is the high-resolution multi-channel imaging system (\citealt{Xu2014}, \citealt{Xiang2016}), including two broad-band interference filter channels (TiO-band and G-band) and three narrow-band Lyot filter channels (H$\alpha$, Ca II 3933\AA\ and He I 10830 \AA). The other is the multi-wavelength spectrograph in visible lines, working at H$\alpha$, Ca II 8542 \AA\ and Fe I 5324\AA \ lines (more information seen in \citealt{Wang2013}). The imaging system and the spectrograph have identical foci and are arranged to be perpendicular to each other as illustrated in Figure~\ref{f:1}. An incoming light is distributed into these two instruments with different percentage using a beam splitter. We employ different kinds of beam splitters to optimize the observation wavelengths for either the simultaneous or the independent operation of these two instruments. In addition, the slit position can be precisely determined in the FOV of the imaging system by an optical calibration technique (\citealt{2016Fu}). The introduction to this calibration process is beyond this paper, but we demonstrate the result in Figure~\ref{f:2}. In this way the imaging system can serve as a slit-jaw recorder for the spectroscopy in despite that the slit is invisible. A field scanner is installed in front of both instruments and comprises two pairs of $K$ mirrors as shown in Figure~\ref{f:1}. One pair of $K$ mirrors move along the optical path to produce the solar image motion in the direction perpendicular to the slit (refer to \citealt{Yang2016} for more details).

In the present work we select a beam splitter which permits 90\% photons to transmit into the spectrograph and 10\% photons entering the imaging system in order to obtain the broad-band filter images in the TiO channel and the high signal-to-noise (SNR) spectra of the H$\alpha$ line (SNR $>$ 100), simultaneously. TiO images are used to detect the solar image motion, including both the scanning and wobble motions. However, at present it is not synchronously recorded with the H$\alpha$ spectra. When serving as a slit-jaw recorder, the camera of the TiO channel is used with 4 $\times$ 4 binning. The recorder rate is then about 15 frames per second with the acquisition time of 60 ms. It is comparable to the value of the H$\alpha$ spectral observation, which is taken typically with 60 ms exposure time and 40 ms readout time, resulting in the acquisition time about 100 ms and the recorder rate about 10 frames per second. The last but not least, the spectral data are taken synchronously with the positioning of the field scanner.
FOV in the TiO channel (slit-jaw) is about 133 $''$ $\times$ 112 $''$ with the pixel sample of about 0.21 $''$. For the spectroscopy observation, the FOV is about 129 $''$ in the slit direction. We usually raster about 110-120 steps with the step size identical with the slit width (0.45 $''$), which corresponds to about 50 $''$- 55 $''$ on the solar surface. If we totally take 5 frames at each scanning position (this parameter is adjustable), it approximately takes about 60 seconds to accomplish such a 110-step rater.

Two-dimensional spectroscopy observations have been carried out for several active regions. In the following, we illustrate our technique to increase the spatial resolution based on two raster observations over the active regions NOAA 12661 (2017 June 7) and NOAA 12671 (2017 August 22). The latter has better seeing conditions than the former.

\section{Methods}
 \label{S:2}
\subsection{Pre-process }
 \label{S:2:1}

\begin{figure}
\centering
\includegraphics[width=14 cm, angle=0]{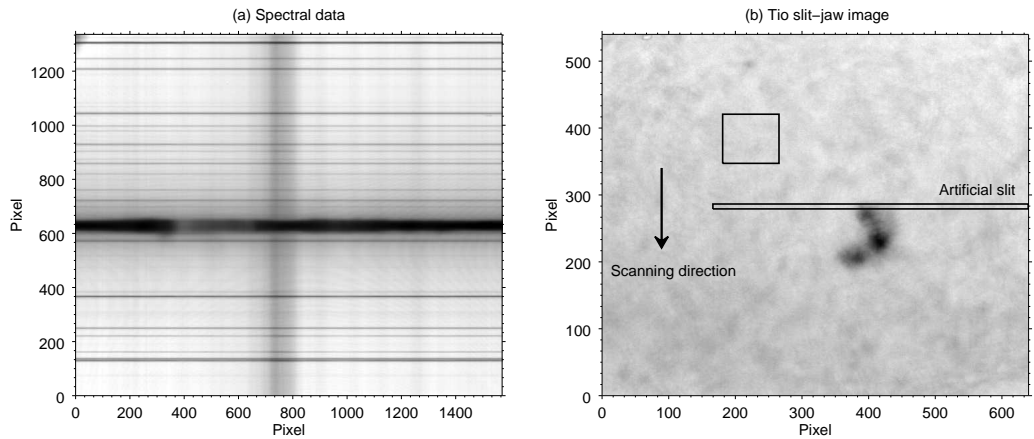}
\caption{(a) One spectrogram after the precise pre-process centering at the H$\alpha$ line. (b) One Tio slit-jaw image taken quasi-simultaneously with the spectrum. An elongated rectangle marks the position of the entrance slit. The arrow indicates the scanning direction. A box outlines the area in which the intensity is integrated and used to evaluate the intensity variation with time.}
\label{f:3}
\end{figure}

Pre-processes of the slit-jaw (TiO) images include the flat-field and dark correction. Unlike \citealt{Johannesson1992}, we do not need to remove a vertical slit from the slit-jaw images. Precisely data pre-reduction of the spectra has been done by \citealt{Cai2017}, which is adapted from the method proposed by \citealt{Wang2013} but specifically applicable for the raster observations. The data reduction comprises the flat-field correction, dark subtraction and distortion correction of spectral lines. In addition, both the wavelength calibration and the continuum intensity modification are seriously carried out for all the spectral data in order to make sure the accuracy of the quantities (e.g., the line-of-sight velocity) retrieved from the spectral data. In other words, each specific spectra line has identical pixel position in all the data to facilitate the following process of 2D map composition. One frame of reduced spectral data and the quasi-simultaneous TiO image are shown in Figure~\ref{f:3}.

\subsection{Calculation the Displacement Vectors}
 \label{S:2:2}
\begin{figure}
\centering
\includegraphics[width=15 cm, angle=0]{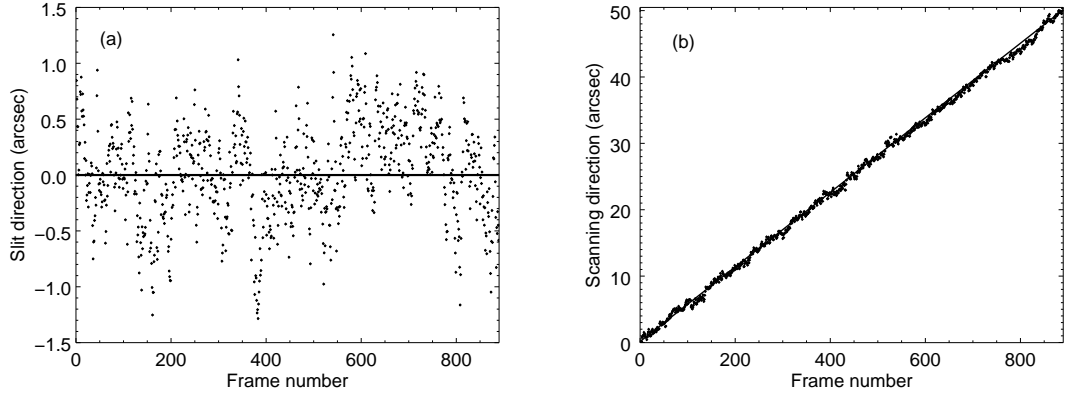}
\caption{Displacement vectors inferred from the TiO slit-jaw images during one scan in both the slit (a) and  scanning (b) directions. A solid line in each panel represents a least-square linear fit.}
\label{f:4}
\end{figure}

As first step, TiO images are used to measure the solar image motion during the raster in both the slit and scan directions. As mentioned above, we usually raster an area about 60$''$ in the scan direction, which can be fully covered by the FOV of the TiO image. In contrast to \citealt{Johannesson1992}, we need not reconstruct a reference slit-jaw image, which is supposed to have a larger FOV to completely cover the observation target throughout the raster. In our case, we choose the middle frame of the TiO images to be the reference frame and calculate the displacement vector of other frames with respect to it using a cross-correction technique. For the  observation of study, sunspot features facilitate this tracking method. The image motion during one scan is shown in Figure~\ref{f:4}. It is found that, in addition to the linear motion in the scan direction represented by a linear fitting, there exists a wobble motion in both the slit and scan directions. The amplitude of this wobble could reach up to 1.3$''$, which is almost 3 times of the step-size (0.45$''$).

\begin{figure}
   \centering
   \includegraphics[width=\textwidth, angle=0]{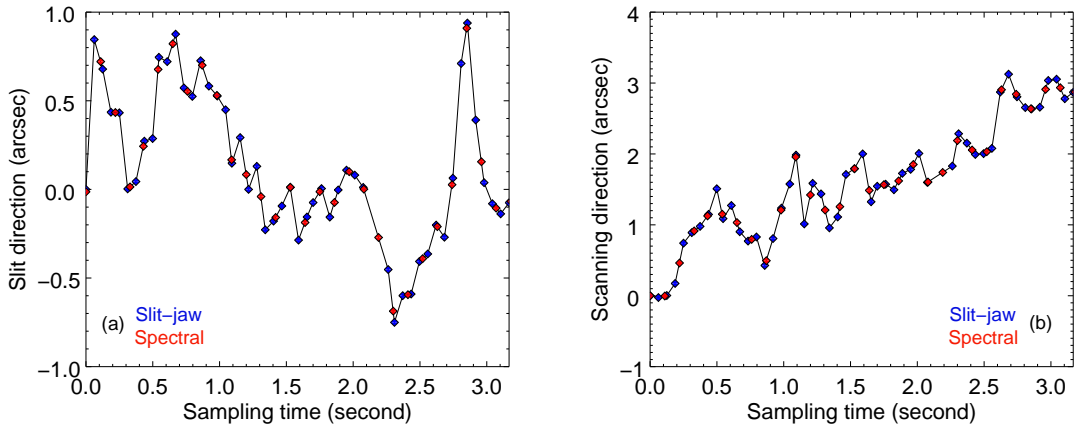}
   \caption{Displacement vectors in both the slit (a) and scanning (b) directions during 3 seconds. Imaging motions at the times when the Ha spectral data (the TiO slit-jaw image) are taken are denoted by red (blue) rhombus.}
   \label{f:5}
\end{figure}

In next, we determine the solar image motion at the time when the spectrogram is observed. It is reminded that the acquisition speed of the TiO slit-jaw image is comparable to that of the H$\alpha$ spectrogram although they are not synchronous. We assume that the variation of the image motion linearly changes with time between two near Tio samples then we implement a linearly interpolation to the displacement vectors detected from the TiO slit-jaw images at the time when the spectrum is taken. Figure~\ref{f:5} shows the displacement vectors of the TiO slit-jaw images and the interpolated vectors of the H$\alpha$ spectral data in about 3 seconds.

\subsection{Correction and Re-arrangement of spectral data}
 \label{S:2:3}
\begin{figure}
   \centering
   \includegraphics[width=\textwidth, angle=0]{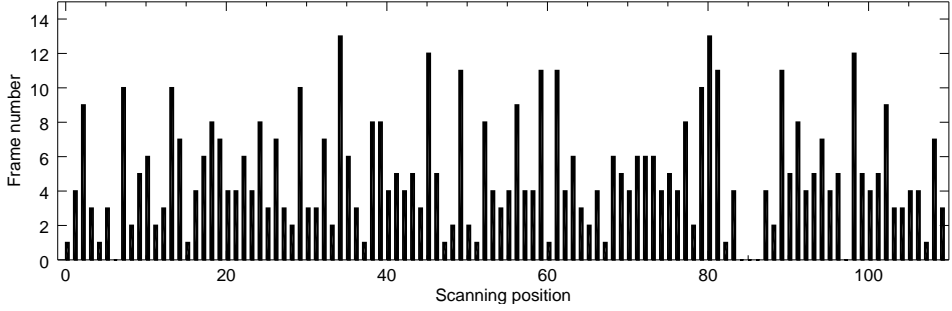}
   \caption{Frame numbers at each scanning position after re-arranged.}
\label{f:6}
\end{figure}

The motion along the slit direction is easily corrected by shifting the spectral data with the corresponding displacement vectors of the slit direction.
However, the motion in the scanning direction needs to be corrected by re-arranging the spectral data according to the displacement vectors of the scanning direction. Due to the wobble motion accompanying with the raster scan, the solar surface can not be evenly scanned by the slit, so that the spatial sample in the scanning direction is not equal to the scan step size. Supposing we initially set the scan step size equal to the slit width (0.45$''$) and consider these spectra, whose displacement vectors are within the range of $n \times 0.45''$ -- $(n+1)\times0.45''$ as the ones taken at the $n$-th scanning position, we find that the frame number at each scanning position is quite different from the initial parameter (i.e., 5 frames). Figure~\ref{f:6} illustrates one example. It is a 110-step raster scan over the active region NOAA 12661 on 2017 June 7 with the step size equal to 0.45$''$ and and temporal resolution of 60 s. It is seen that the frame number of the spectral data re-arranged into one scanning position is various. The maximum frame number is 13 while the minimum is 0, which indicates that some areas are multi-time scanned but some are not scanned at all.

\subsection{Composing into a Two-dimensional Map }
 \label{S:2:4}
At last, we combine all the spectra into the 2D maps. We have different strategies of the composition for two situations mentioned above, i.e., the multiple-frame and 0-frame cases at each scanning position.

\emph{Case1}: When there are multiple frames harbored at one scanning position, we take the averaged frame of the multiples with the assumption that all of them basically correspond to the same position of the solar surface within a certain spatial and temporal resolutions. Actually we also apply the frame-selection methods with different criteria, such as, to select the spectrum that is taken when the quasi-simultaneous TiO slit-jaw image has the strongest image contrast or to pick up the spectrum which shows the deepest specific line depth. However we do not find any distinct improvement in the quality of composed image, we therefore use the averaged frame in practice.

\begin{figure}
   \centering
   \includegraphics[width=\textwidth, angle=0]{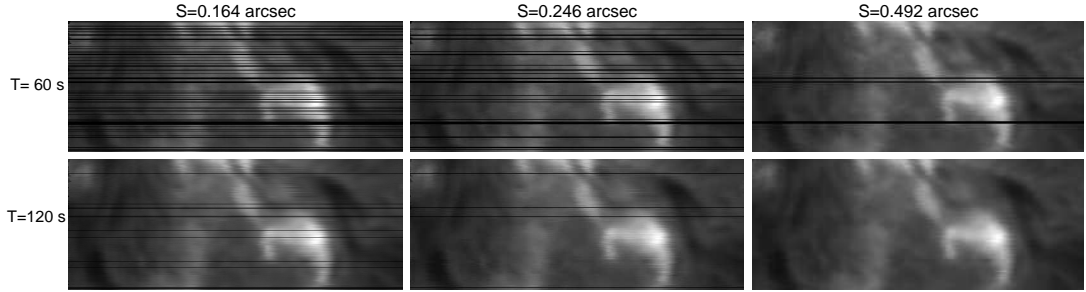}
   \caption{Two-dimensional intensity map of the active region NOAA 12661, composed at the H$\alpha$ line center with the different spatial samples and temporal resolutions.}
\label{f:7}
\end{figure}

\emph{Case2}: If there is 0 frame at a scanning position, we usually take the cubic-interpolation into the neighboring values when the 2D composed image is accomplished. However when such a 0-frame case takes place at many scan steps, we need to consider to a new spatial sample and temporal resolution to re-arrange the spectral data in space, instead of the fixed step size. As shown in Figure~\ref{f:7} (from left to right), we consider different spatial samples of 0.164 $''$, 0.246 $''$ and 0.492 $''$ with 60 s temporal resolution, respectively, to re-arrange the sequentially observed spectral data into the right position. When the spatial sample is set as small as 0.164 $''$, which is about two times of the pixel sample along the slit of the spectral data (0.082 $''$/pixel), it is found that several scanning positions show the 0-frame case displayed by the black lines in the composed monochromatic intensity map. However when we decrease the spatial sample into a value of 0.492 $''$, the black lines become much few. Alternatively, we can increase the initial acquisition frame number at each scan step or combine several successive scans into one. Of course it is inevitable to decrease the temporal resolution of the raster scan. Also as shown in Figure~\ref{f:7}, we compose a 2D intensity map using the spectral data taken in two successive scans (i.e., in the bottom row). In this case, it is equal to taking $one$ dense scan over the FOV of 50 $''$, meanwhile the time consumption (i.e., the time evolution) is correspondingly extended to about 120 s.

In short, both the spatial sample and the temporal resolution have to be reasonably taken into account to re-arrange the spectra in order to eventually compose a 2D map to meet different needs of various scientific studies.

\section{Results and Evaluation}
\label{S:3}
\begin{figure}
   \centering
   \includegraphics[width=\textwidth, angle=0]{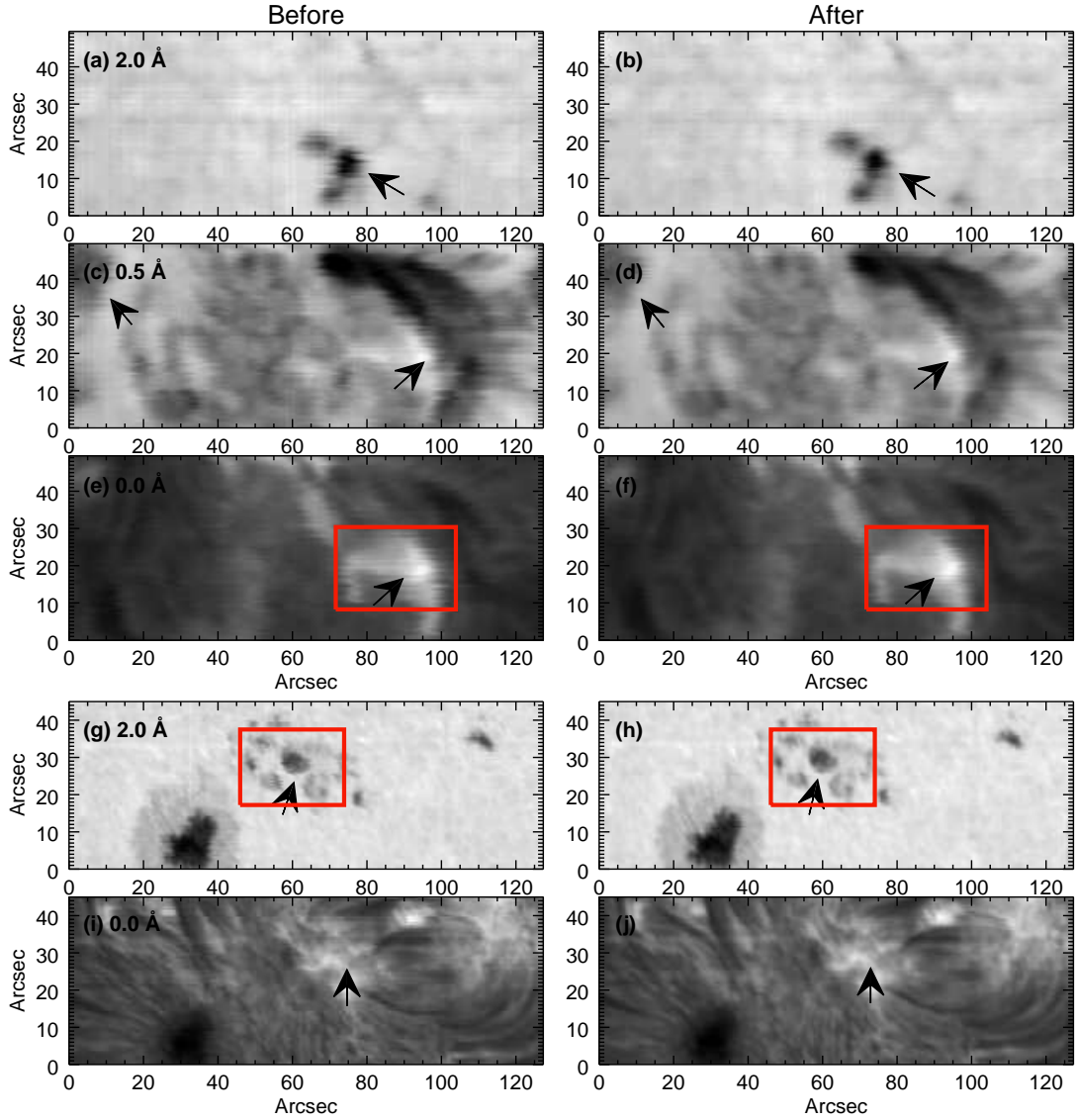}
   \caption{Two-dimensional composed intensity map at different wavelengths. (a-f) A 110-step raster over the active region NOAA 12661. (g-j) A 100-step raster over the active region NOAA 12671. Two-dimensional maps are composed and shown at different wavelength around the H$\alpha$ line as the notes indicate (2.0 \AA, 0.5 \AA \ and 0 \AA.). Red rectangles outline the sub-regions in which the second-order finite differences are calculated and compared.}
   \label{f:8}
   \end{figure}

We compare the 2D composed intensity maps before and after solar motion correction in Figure~\ref{f:8}. Two times of observation are exhibited. Figure~\ref{f:8}(a)-(f) show a 110-step raster over the active region NOAA 12661 on 2017 June 7, while Figure~\ref{f:8}(g)-(j) show a 100-step raster over NOAA 12671 on 2017 August 22. The left columns show the results which are directly composed by the spectra taken sequentially, and the right columns show the ones composed by the spectra re-arranged in scan direction and corrected in the slit direction. The spatial sample is considered to be the same for both data sets (i.e., equal to the initial scan step size 0.45 $''$). It is clearly seen that some solar structures in the original map have much more burrs than these in the map after the motion correction, particularly in the regions pointed by the arrows, such as the sunspots with strong intensity contrast in the photosphere (e.g., Figure~\ref{f:8}(a), (b), (g) and (h)) and a filament located at the adjacent quiet region (e.g., Figure~\ref{f:8} (c) and (d)). We also find that although the seeing in the second data is much better than the first, the burrs are still clearly present around the sunspots.
It means the most of these burrs are caused by the malposition and the drift of the slit position during the scanning.

The above burrs can result in the intensity fluctuation and even the structure deformation in 2D map, therefore we use two methods to quantitatively evaluate the composed images. Firstly we pick up a sub-region, in which sunspots or bright structures are present. Then we calculate the second-order finite differences in this region using every three neighboring pixels. The $rms$ of the second-order finite difference reflects the intensity fluctuation. For example, in the first observation we focus on a bright plage seen in the monochromatic image of the H$\alpha$ line center (Figure~\ref{f:8}(e) and (f)). The $rms$ is calculated to be about 0.021 based the original 2D map and 0.015 based on the corrected image. In the second observation which has a better seeing condition, we focus on a sunspot area seen in the monochromatic image at continuum (Figure~\ref{f:8}(g) and (h)). The $rms$ is about 0.020 before the motion correction and becomes to be 0.016 after. Both results show that almost 1/4 intensity fluctuation is caused by the burrs and eliminated after the motion correction.Secondly we compare the similarity between the 2D intensity image composed at continuum (seen Figure~\ref{f:8}(a-b) and (g-h)) and the TiO slit-jaw images. This comparison is in terms of the Structural Similarity Index Measurement (SSIM). The TiO slit-jaw images need to be firstly resized to the same spatial resolution as the composed maps. Specifically, for the first observation, the SSIM between the TiO slit-jaw image and Figure~\ref{f:8}(a) is about 0.85. After the motion correction (e.g.,Figure~\ref{f:8}(b)), the SSIM is about 0.91. For the second observation, the SSIM increases from about 0.90 to 0.94 after the motion correction. In brief, the quantitative evaluations resulted from two methods show the quality of the raster images is improved after the motion correction. It is also found that given a certain spatial sample and temporal resolution, the spatial resolution of the composed 2D map could be close to that of the slit-jaw image.

\begin{figure}
   \centering
   \includegraphics[width=\textwidth, angle=0]{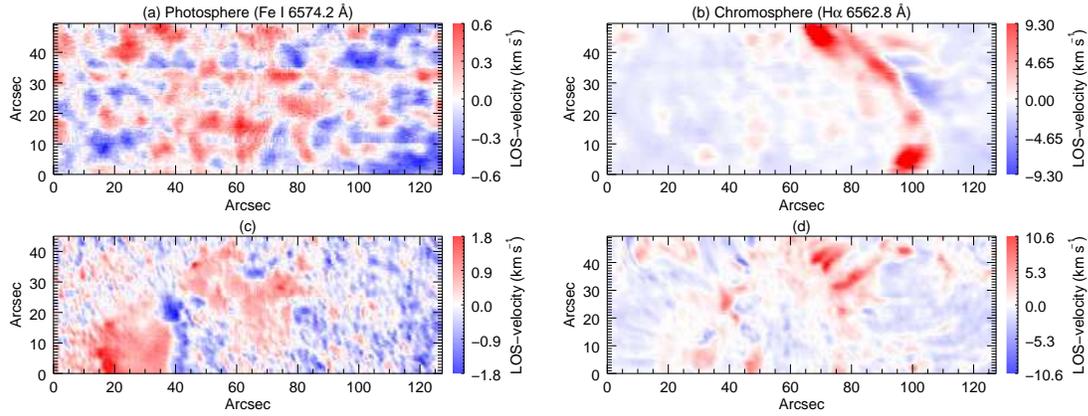}
   \caption{Two-dimensional Doppler velocity maps retrieved from the scanning spectra over the active region NOAA\ 12661 (a-b) and NOAA\ 12671(c-d).}
   \label{f:9}
\end{figure}

Besides the measured quantities like the line intensity, we can also obtain the 2D map of inferred quantities like the line-of-sight (LOS) velocity. The LOS velocity is derived using the center-of-gravity method from the specific line profile. Meanwhile the 32\AA-wavelength range in the H$\alpha$ band allows us to observe the photospheric and chromospheric lines simultaneously. In addition, the telluric lines can also be used as wavelength reference to calculate the absolute LOS velocity. As shown in Figure~\ref{f:9}, we display the LOS velocities of two active regions in the photosphere in the panel (a) and (c). It is retrieved from the photospheric Fe {\sc I} (6574.2\,{\AA}) line.
Besides the granular pattern of the velocity, the large-scale Evershed flows around sunspots are clearly present along the penumbra filaments. It is revealed that the speed varies from around 1 km/s at the border between the umbra and the penumbra to a maximum of 1.8 km/s in the middle of the penumbra and falls off to zero at the outer edge of the penumbra. This spatial distribution is in agreement with the previous observation (\citealt{Rimmele2006}).
Panel (b) and (d) are the LOS velocities in the chromosphere inferred from the H$\alpha$ line. Although the Evershed flow around sunspot is not very distinct here, our other observation data clearly show a reversal flow present in the chromosphere (\citealt{Xu2010}). In addition, large scale velocities are found around the filament structure, while fast down-flows are detected at the footprints of the arch filament system.

\section{Conclusion and Discussion}
\label{S:4}

It is allowed to employ the multi-channel high resolution imaging system and the multi-wavelength spectrograph co-temporally, co-spatially and co-foci at the NVST. The slit position can be precisely determined in the FOV of the imaging system, which makes it possible to use the imaging system as a slit-jaw recorder to monitor the solar image motion during the scanning, even there is no visible slit in its FOV.

Taking advantage of this convenience, we detect the spectra displacement vectors of the solar image motion during the spatial scanning from the broad-band imaging observations (e.g., the TiO-band). The displacement vectors includes both the scanning and wobble motions. Without any image stabilizer, the random motion of the solar image under studying could instantaneously reach up to 1.3 arc sec or more, which is almost 3 times larger than the spatial sample (typically equal to the slit width 0.45$''$) and severely decreases the spatial resolution of the 2D spectroscopy observations.

As next step, these displacement values need to be linearly interpolated to get the values when the spectral data (e.g., the H$\alpha$) are quasi-simultaneously taken. For the present, the slit-jaw images and the spectral data are not synchronously acquired, but the acquisition rates are comparable. After that, the series of spectral data originally taken in sequence with time are corrected (in slit direction) and re-arranged (in scanning direction) in space. Two techniques are used to evaluate the resulting quality of newly composed 2D map. The spatial resolution is fairly improved and closes to the one of the slit-jaw images.

However in the corrected raster 2D maps there still exist a small amount of burrs, i.e., the discontinuity of the solar structure border which is supposed to be smooth. The reason could be that the acquisition speeds of spectral data and slit-jaw images are not strictly synchronized. The interpolated displacement vectors of the spectral data might be different with the real case. Synchronous observations need to be carefully considered in the next step. In addition, the cross-correlation is used to calculate the displacement vectors of the Tio slit-jaw images in this work, the premise of this algorithm can be achieved successfully is the existence of obvious structures (e.g., sunspots or pores) inside the FOV of the slit-jaw images. If there are no any these structures, the effect of correction would much less prominent since the image cross-correlation algorithm may not work very well. There are two solutions for this case. On the one hand, we expect the seeing is so good that the granule structures can be used to do the image cross-correlation. On the other hand, we can employ the image channel of the He I 10830 or H$\alpha$ to be the slit-jaw recorder since it is possible that more ¡°apparent structures¡± are presented in the chromosphere than in the photosphere.

From this work we also realize the necessity of the image stabilizer in order to reduce the image random fluctuation during scanning, which is presently quite large compared with the spatial sample. Otherwise, the enormous amount of data are needed when one wants to achieve a high spatial resolution of the 2D spectroscopy by using the technique under study. It is not very practical for the high temporal resolution requirement of an active region which continuously evolves with time. Given the present parameter setting (i.e., the spatial sample is of 0.45 $''$. Five frames are acquired at each slit position. The exposure time of H$\alpha$ line is about 60 ms), we can accomplish a 2D map with 60$''$ in the scanning direction within 60 s and achieve the spatial resolution about 2 arc sec.

\normalem
\begin{acknowledgements}
We are appreciated for all the help from the colleagues in the NVST team. This work is supported by the National Natural Science Foundation of China (NSFC) under grant numbers 11773072 and 11573012, 11473064.
\end{acknowledgements}

\bibliographystyle{raa}
\bibliography{ms0231}

\end{document}